# Investigation of Intrinsic Properties of High-Quality Fiber Fabry-Perot Resonators

GERMAIN BOURCIER,[1,*] STEPHANE BALAC,[2] SAFIA MOHAND-OUSAID,[1] JULIEN LUMEAU,[3] ANTONIN MOREAU,[3] VINCENT CROZATIER,[4] PIERRE SILLARD,[5] MARIANNE BIGOT,[5] LAURENT BIGOT,[6] OLIVIER LLOPIS,[1] AND ARNAUD FERNANDEZ,[1]

[1] LAAS-CNRS, Université de Toulouse, 31400 Toulouse, France.
[2] IRMAR, Univ. Rennes, CNRS, IRMAR, F-35000, Rennes, France.
[3] Aix Marseille Univ, CNRS, Centrale Med, Institut Fresnel, Marseille, France.
[4] Thales Research and Technology, Palaiseau, France.
[5] Prysmian, Parc des Industries Artois Flandres, 644 boulevard Est, Billy Berclau, 62092 Haisnes Cedex, France
[6] University of Lille, CNRS, UMR 8523-PhLAM Physique des Lasers Atomes et Molecules, F-59000, Lille, France

*gbourcier@laas.fr

**Fiber Fabry-Perot (FFP) resonators of few centimeters are optimized as a function of the reflectivity of the mirrors and the dimensions of the intra cavity waveguide. Loaded quality factor in excess of $10^9$, with an optimum of $\sim 4 \times 10^9$, together with intrinsic quality factor larger than $10^{10}$ and intrinsic finesse in the range of $10^5$ have been measured. An application to the stabilization of laser frequency fluctuations is presented.**

Optical resonators are key components in a wide range of applications [1]-[4], from optical communications and frequency references to quantum sensing and biosensors. Their assessment predominantly hinges on two partially interrelated parameters: finesse, *F*, and quality factor, *Q*, conditioned by intracavity losses, coupling, and resonator dimensions. *F* serves as an indicator of the technology's inherent quality, regardless of resonator dimensions, while *Q* governs system performance. High values of *F* and *Q* are essential. In the linear regime, for example, such resonators enable efficient stabilization of optical [5] and microwave sources, notably in opto-electronic oscillators (OEOs) [6]. They also enhance or enable a variety of nonlinear optical processes [7]. Nevertheless, *F* and *Q* represent merely a fraction of the points to be considered. Optimizing the integration within the system, especially controlling resonator coupling, is of crucial importance. Numerous resonator technologies have been proposed, each presenting unique advantages and disadvantages. Ultra-Low Expansion (ULE) Fabry-Perot resonators exhibit exceptionally high *Q* and *F* ($10^9 < Q < 10^{11}$ and $F > 10^5$ [8]). Nonetheless, they tend to be bulky and are mainly used in the linear regime to realize ultra-stable lasers. Whispering Gallery Mode (WGM) resonators also boast remarkably high *Q* factors up to $10^8$ [6], [9], and *F* exceeding $10^6$ [9]. However, they encounter challenges in manufacturing and achieving reproducible coupling control. Integrated ring resonators leverage cutting-edge advancements in silicon photonics, offering inherent *Q* factors surpassing $10^8$ [7]. Fiber ring cavities extending over several meters similarly provide high *Q* factors ($\sim 10^{10}$) [10] along with reliable coupling. However, the use of a fairly long fiber length makes the resonator sensitive to thermal and mechanical disturbances linked to the fluctuation of the optical path. FFP resonators, comprising a few centimeters of optical fiber bounded by two dielectric mirrors, bridge the divide between fiber loops and integrated resonators. They combine the benefits of both technologies, presenting flexibility, reproducibility, robustness, and ease of coupling thanks to standard FC/PC connectors, facilitating integration into fiber-based systems. Moreover, the excellent transparency of optical fibers, coupled with the diverse dispersive and nonlinear properties of these devices, offers an unmatched array of resonant waveguides not achievable by other optical resonator technologies developed thus far. Since their first appearance in 2009 [11], FFP resonators have gained significant importance [12], [13].

The investigations presented in this paper aim to quantify intra-cavity losses induced by mirror diffraction, which fundamentally limit the optimal finesse of FFP resonator at maximum storage efficiency. This study was conducted using optical fibers with varying effective mode areas ($A_{eff}$) and mirrors with different reflectivities (*R*). We experimentally and numerically demonstrate the influence of $A_{eff}$ on *F*. In addition, our results show that intrinsic quality factors ($Q_{int}$) and finesse ($F_{int}$) exceeding $10^{10}$ and $10^5$, respectively, can be achieved. Finally, we highlight a practical application that benefits from such high-quality-factor resonators: the reduction of laser frequency noise.

Our FFPs consist of a $L \cong 7$ cm sample of optical fiber resulting in a free spectral range $FSR = c/(2n_gL)$ of 1.45 GHz, delimited by two few microns thick dielectric mirrors deposited via plasma assisted electron beam deposition (Bühler Leybold Optics SYRUSpro). Each mirror is composed of a stack of λ/4 layers of $Nb_2O_5$ ($n_H$=2.18) and $SiO_2$ ($n_L$=1.47). *R* is monitored in-situ with a reflectance measurement of the multilayer stack deposited on a control sample. It allows feedback with the model all along the process, which leads to an accuracy on the reflectance close to 0.01%. The choice of the 7 cm length corresponds to almost the smaller size we can afford to get a single pass mirror deposition on both fiber ends (fiber in U shape during deposition). It results in a resonator with microwave range FSR, which should be easy to mechanically and thermally stabilize thanks to its relatively small size.

In this study, we focus on the finesse $F = FSR/\delta f_{1/2}$ and quality factor $Q = \nu_0/\delta f_{1/2}$, with $\delta f_{1/2}$ the full width at half maximum (FWHM) of the FFP resonance, measured at the optical frequency $\nu_0$ =193.4 THz. To determine the best achievable finesse for a given FFP design, we fabricated several groups of up to 15 resonators, all with identical mirrors at both ends. Each group differs by either the mirror reflectivity or the effective area of the fiber. As illustrated in Fig. 1, the resonators were characterized using a RF spectroscopy technique. A narrow-linewidth external cavity laser (ECL) diode (RIO Planex @1.55 µm) was locked to a resonance mode, while adjacent side modes were probed via optical beating using a Mach-Zehnder modulator (MZM), a vector network analyzer (VNA) and a photodetector (PD$_1$) [14]. As a result, the amplitude and phase of the measured S$_{21}$ parameter provide an exact replica, in the microwave domain, of the optical complex transfer function of the FFP (see inset of Fig. 1).

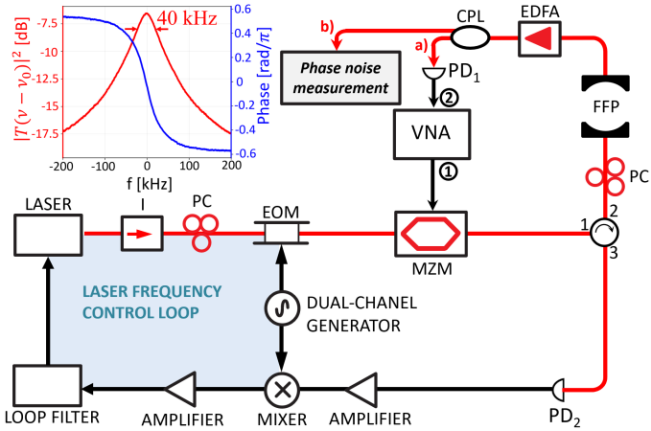

Fig. 1. Experimental setup used for both: a) acquisition of the FFP transfer function and the recovery of the parameters $a$ and $r$; and b) laser frequency noise measurement. The optical fibers (red line) are single-mode fibers.

The parameters of the FFP such as the intra-cavity losses $a \in [0,1]$ (attenuation factor) and mirrors reflectivity $r \in [0,1]$ encountered by the electromagnetic field amplitude are extracted by using the general model of Yariv based on a transfer matrix formalism to describe the resonator [15]. $a = a_p a_d^2 a_a^2$ with $a_p$ the propagation losses in the fiber, $a_d$ and $a_a$ respectively the diffraction and the absorption losses on the mirrors for a cavity roundtrip in amplitude, although the propagation losses are negligible compared to the two other terms (see further discussion). $Q_{int}$ is obtained by artificially setting the reflectivity to 100 % ($r$ = 1) which results in a totally isolated (uncoupled) resonator. According to equations (1) and (2) given below, $\delta f_{1/2}$ and the maximum transmission at resonance $T_{max}$ issued from S$_{21}$ measurement are the pivotal values that help to recover in a straightforward fashion $a$ and $r$ through the relation $R = r^2$ where $R$ is the power reflectivity.

$$\delta f_{1/2} = \frac{c}{2\pi n_g L} Arccos\left(\frac{2aR}{1+(aR)^2}\right) \quad (1)$$

$$T_{max} = \frac{a(1-R)^2}{(1-aR)^2} \quad (2)$$

We proceeded to $\delta f_{1/2}$ measurement by using an EDFA (Fig. 1) in order to recover the product $aR$ with a sufficiently high definition. As expressed in eq. (2), $T_{max}$ allows for the decoupling of the two variables $a$ and $R$. It is however measured without EDFA in order to remove uneven saturation of the amplifier thus hindering a precise measure of $T_{max}$. The $R$ computation with this technique allows a further improvement of the precision compared to the direct measurement performed during the mirror processing. An accuracy of 0.001% on $R$ is obtained in most experimental cases.

Previous studies by Jia *et al.* [16] have shown that diffraction losses located in the Bragg dielectric mirrors are significant and that these losses are considerably reduced by using FFP resonator based on multimode fiber (MMF) due to a larger modal area. In order to bring a quantitative comprehension of this assumption, we investigated the impact of the fiber $A_{eff}$ on the resonator's performance. For this purpose, we have characterized four sets of resonators, all with mirrors reflectivity of $R$ = 99.86 %, but each made from fibers with different $A_{eff}$ for the fundamental mode (LP$_{01}$), ranging from 13 to 198 µm². For resonators based on few mode fibers (FMF) and MMF, we employed specific fiber mode adapters to match the LP$_{01}$ and prevent the excitation of higher-order modes.

We have plotted in Fig. 2 the highest measured finesse for each set of resonators as a function of the LP$_{01}$ mode effective area (blue points). It can be observed that $F$ increases rapidly with $A_{eff}$ and reaches a plateau around ~1800 for FFP samples based on MMF, corresponding to intra-cavity losses of $a_{dB}$ =3.12×10$^{-3}$ dB ($a$ = 0.9996). It is worth noting that the same FFP, limited solely by the intrinsic propagation losses of low-loss silica fiber ($a_{p,dB}$ =2.8×10$^{-5}$ dB, computed from 0.2×10$^{-5}$ dB/cm in the guide) and mirror absorption estimated at 200 ppm ($a_{a,dB}$ =8.68×10$^{-4}$ dB) would theoretically yield a finesse of 1955. This value remains lower than the ideal finesse of a resonator operating at transparency ($a$ = 1), for which the finesse is given by $F = \pi\sqrt{R}/(1-R)$ yielding a value of 2242. Due to the low linear losses, $F$ can be considered independent of the FFP length. In contrast, $Q = (2n_g\nu_0/c)FL$ is proportional to the length. For comparison purposes, all the resonators were designed with identical lengths.

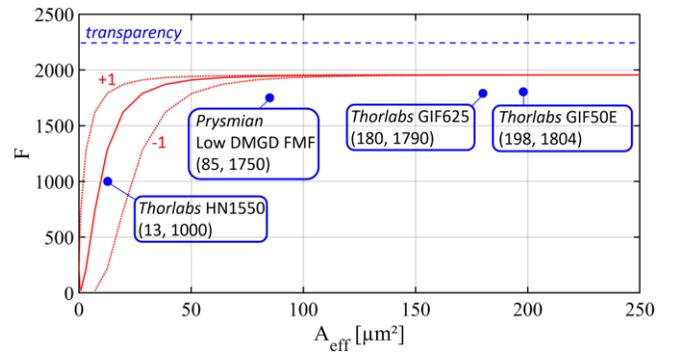

Fig. 2. Experimental measurements of the finesse as a function of the effective area of the resonator's fiber mode (blue dots), along with numerical calculations of the finesse accounting for diffraction losses at the mirrors (red line). The impact of a ±1 µm uncertainty in the initial beam waist value is represented by the red dashed lines.

We carried out numerical investigations based on Gaussian mode analysis, approximating the fundamental mode in optical fibers and treating the fiber's effective area as a known parameter. The beam waist $w_0$, defined by the Gaussian's standard deviation, corresponds to the radius at $1/e^2$, such that $A_{eff} = \pi w_0^2$. As illustrated in Fig. 3.(a), the intra-cavity fundamental mode incident on the mirror ($\vec{E_0}$) is decomposed as a plane waves expansion using a bidimensional Fourier transform with respect to the spatial coordinates (x, y) in the plane orthogonal to the propagation axis chosen as the z-axis. The Fourier variables (spatial frequencies) $k_x$ and $k_y$ physically correspond to the wave vector coordinates within this plane. Since the magnitude of the wave vector is known, the longitudinal component $k_z$, corresponding to the direction of propagation and normal to the mirror surface, can be deduced. This component $k_z$ defines the angle of incidence of each plane wave onto the Bragg mirror. Thus, each plane wave in the fundamental mode expansion is characterized by an incidence angle $\theta_i$ and a complex-valued amplitude $E_0(k_x, k_y)$. To determine each plane wave's reflection coefficient, we have calculated the global Fresnel coefficients across 10 pairs of λ/4 thin-film layers described above by using a recursive method [17]. The computed reflection coefficients are applied in Fourier space, and an inverse Fourier transform yields the reflected spatial mode distribution ($\vec{E_r}$).

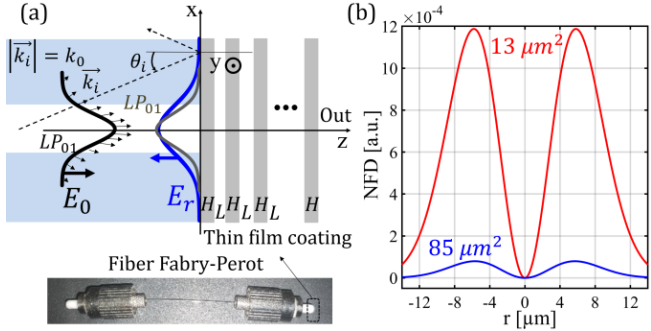

Fig. 3. (a) Illustration of the intra-cavity fundamental mode decomposed into a plane waves expansion incident to the right thin-film coating of the FFP resonator. (b) Numerical results showing the normalized field difference obtained with two distinct incident $A_{eff}$ of 13 and 85 µm² with a mirror reflectivity of 99.86 %.

By comparing the reflected and incident mode powers at the mirror, we determine the mirror's power reflectivity for a given Gaussian mode and thus for a specific $A_{eff}$. The diffraction occurring at the mirror causes a broadening of the reflected Gaussian mode, which increases as $A_{eff}$ decreases as illustrated in Fig. 3.(b) by the normalized field difference (NFD) defined as: $NFD = |E_r|^2_{norm} - |E_0|^2_{norm}$. Since the reflected mode consistently exhibits an $A_{eff}$ larger than the LP$_{01}$ mode effective area supported by the fiber, modal mismatch arises leading to intra-cavity losses (parameter $a_d$) that we calculate using the overlap integral. Hence, for each specific $A_{eff}$, the corresponding finesse of the FFP resonator is deduced from this approach and shown in Fig. 2 (red curve). This indicates that diffraction losses dominate for $A_{eff} < 80\ \mu m^2$. Above this threshold, the FFP finesse remains constant and is limited only by fiber background losses and mirror absorption. The experimental measurements closely match the numerical results. However, the lower measured ceiling value reveals additional physical limitations due to imperfect polished fiber ends (surface roughness [16], apex mismatch [18]) and FC/PC connector imperfections (eccentricity, bore size tolerance).

$$\text{PEF} = \frac{1-R}{(1-aR)^2} \quad (3)$$

In order to achieve the best possible finesse at maximum power enhancement factor (PEF) [10], [14], [15], defined by relation (3) as the ratio between the intra-cavity and injected power, we focused on resonators with the largest effective area. We thus fabricated and characterized three sets of 15 resonators with the same MMF (*Thorlabs* GIF50E) with an $A_{eff}$ of 198 µm², combined with mirrors having reflectivities of 99.93, 99.96 and 99.99 % respectively. Fig. 4.(a) presents the best measured $Q$ factor and $F$ as functions of the deposited mirror reflectivity. The experimental data show that both $Q$ and $F$ increase with reflectivity. Indeed, assuming $aR \cong 1$, a 2nd order Taylor expansion of relation (1) yields a simplified expression for $\delta f_{1/2}$ (eq. (4)) indicating that $F \approx \pi/(1-aR)$ and $Q = \nu_0 F/FSR$ both depend on $1/(1-aR)$.

$$\delta f_{1/2} = \frac{c}{2\pi n_g L} \frac{(1-aR)}{\sqrt{aR}} \quad (4)$$

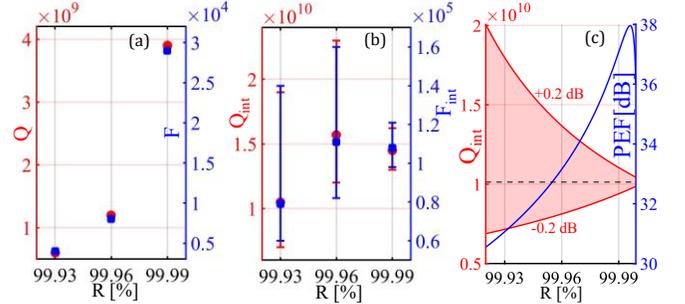

Fig. 4. Quality factor and finesse versus mirror reflectivity. (a) Measured values. (b) Deduced intrinsic values. The ±0.2 dB uncertainty in $T_{max}$ measurement using the VNA (Fig. 1) affects the accuracy of intra-cavity loss retrieval and thus the calculation of intrinsic $Q_{int}$ and $F_{int}$. (c) Graphical illustration of constant $Q_{int} = 1.1 \times 10^{10}$ (black dashed line) versus reflectivity $R$ for a FFP with fixed intra-cavity loss $a = 3.5 \times 10^{-4}$ dB and its uncertainty (red area) arising from a ±0.2 dB uncertainty in the experimental measurement of $T_{max}$.

As shown in Fig. 4.(a), a maximum $Q$ factor of $3.9 \times 10^9$, corresponding to a finesse of 29400, is measured with R=99.99 %. Fig. 4.(b) shows the deduced intrinsic values of $Q_{int} = \frac{2\pi \nu_0 L}{v_g(1-a)}$ [14], and $F_{int}$ of the resonators, which isolate the intra-cavity losses defined by $a$. The uncertainty bars reflect the sensitivity of $a$ to variations in $T_{max}$, calculated from (2) as follows,

$$\frac{da}{dT_{max}} = \frac{a}{T_{max}} \frac{1-aR}{1+aR} \quad (5)$$

From eq. (5) the uncertainty $\Delta a$ due to ±0.2 dB uncertainty in $T_{max}$ can be expressed as,

$$\Delta a = \pm \varepsilon a \frac{\pi}{2F-\pi} \quad (6)$$

with $\varepsilon \simeq 0.047$ and $F = \pi/(1-aR)$ as $F$ (or equivalently $aR$) is accurately measured regardless of the value of $R$. Fig. 4.(c), illustrates that $\Delta a$ decreases monotonically as $R$ approaches 1. This is understood by considering the well-known relation $Q^{-1} = Q_{ext}^{-1} + Q_{int}^{-1}$. In the over-coupling regime ($r < a$), mirror coupling losses dominate intra-cavity losses ($Q_{ext} \ll Q_{int}$), so the loaded $Q$ factor is governed by $Q_{ext}$ with $Q_{ext} = \frac{2\pi v_0 L}{v_g(1-r)}$ [14], resulting in large uncertainties in $Q_{int}$ evaluation. Conversely, in the under-coupling regime the loaded $Q$ is dominated by $Q_{int}$, improving confidence in the calculation of $Q_{int}$ for the highest deposited reflectivity of 99.99 % as analytically confirmed in Fig. 4.(c).

Our FFP resonators based on the combination of MMF and $Nb_2O_5/SiO_2$ thin-film mirrors with 99.99 % reflectivity reach a record intrinsic quality factor $Q_{int}$ =1.4×10$^{10}$ and intrinsic finesse $F_{int}$ =1.1×10$^5$ by minimizing the intra-cavity losses down to the fiber background losses, $a_{dB} = 2.8 \times 10^{-4}$ dB. This record was achieved under an optimized coupling regime with respect to $a$, where the PEF closely approaches its maximum achievable value (Fig. 4.(c)) for a transmission at resonance $T_{max} \approx$ -6 dB. The measured optical transmission and characteristic phase transition for this resonator are shown in the inset of Fig. 1.

The measured performances of our FFP resonator elevates this technology to a new level, enabling state-of-the-art operation and paving the way for novel applications. One direct use of high-$Q$ resonators is the stabilization of laser frequency fluctuations. To this end, we employ a Pound Drever Hall (PDH) electronic feedback system on previously presented ECL. Using the same configuration as before, the ECL is locked on the high-$Q$ mini-resonator. The optical signal at the output port of the FFP resonator is then routed to an optical frequency noise analysis system (Fig. 1, branch b). This system includes a self-heterodyne frequency discriminator consisting of a 2 km fiber delay line, an acousto-optic modulator and a low phase noise 80 MHz RF source [19]. The system is enclosed in a metal housing and protected from vibrations, which allows to reach a frequency noise floor close to the system noise floor measured in [19] without fiber spool, which is below 10$^{-2}$ Hz$^2$/Hz above 2 kHz offset and close to 100 Hz$^2$/Hz at 10 Hz offset.

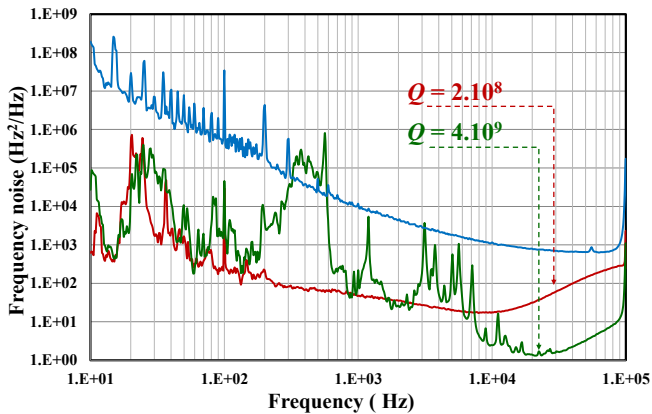

Fig. 5. Frequency noise of the ECL stabilized on a 7 cm long MMF resonator with $Q$ = 4×10$^9$ (green) and on a 2 cm long MMF resonator sealed in a ceramic ferrule with $Q$ = 2×10$^8$ (red) is compared to the free-running ECL laser (blue).

As shown in Fig. 5, the ECL stabilized on the 7 cm long MMF resonator ($F$ = 30×10$^3$, $Q$ = 4×10$^9$) exhibits higher frequency noise level at frequency offsets below 10 kHz compared to its stabilization on a 2 cm long ferrule MMF resonator ($F$ = 5×10$^3$, $Q$ = 2×10$^8$) [20], indicating increased sensitivity to mechanical perturbations of the unprotected fiber. Conversely, despite its lower $Q$, the ferrule mini-resonator shows superior robustness with vibrations impacting it only at very low frequencies (f < 100 Hz). From the frequency noise data, the integrated linewidth of the laser stabilized on the ferrule FFP resonator was computed following the approach described in [5]. We have calculated the laser phase noise and plotted the integrated phase noise from high offset frequencies to low offsets. When this integral reach 1 rad$^2$, the noise sidebands power equals the carrier power. Using this formalism, the stabilized laser linewidth is 240 Hz with 2 cm long ferrule MMF resonator. In the case of the 7 cm resonator, the vibration peaks significantly increase the linewidth, though better phase noise is observed at 20 kHz offset, suggesting that this system could be highly effective if near-carrier noise could be mitigated. Future efforts will focus on improving the mechanical and thermal stability of this FFP.

We have conducted measurements revealing remarkably high optical $Q$ factors and finesse in FFP resonators with a state-of-the-art intrinsic finesse of 1.1×10$^5$, presenting promising opportunities of applications. Our findings underscore the substantial influence of the fiber effective area on the performance of these resonators, emphasizing the imperative need for adapting mirror reflectivity to the fiber's effective area in order to maximize the PEF. Additionally, we provide an example demonstrating laser stabilization using two of these resonators with different fiber protective sleeve.

**Funding.** The present research was supported by the French Ministry of Armed Forces – Agence de l'Innovation Défense (AID) and the national space center (CNES).

**Disclosures.** The authors declare no conflicts of interest.

**Data availability**. Data underlying the results presented in this paper are not publicly available at this time but may be obtained from the authors upon reasonable request.